\documentclass[11pt]{article}
\usepackage[english]{babel}
\usepackage{amsmath,amsthm,amssymb,mathtools}
\usepackage[letterpaper,margin=1in]{geometry}
\usepackage{xcolor}
\usepackage{tikz}
\usepackage{algorithm}
\usepackage[noend]{algpseudocode}
\usepackage[pagebackref,colorlinks=true]{hyperref}
\usepackage[capitalize]{cleveref}
\usepackage[T1]{fontenc}

\usepackage{interval}
\intervalconfig{soft open fences}

\hypersetup{
  linkcolor=blue!55!black,
  citecolor=green!40!black,
  urlcolor=blue!65!black,
  pdftitle={Optimal Diamond Distance Estimation for Unitary Channels},
  pdfauthor={Jinge Bao and Qisheng Wang}
}

\newtheorem{theorem}{Theorem}[section]
\newtheorem{lemma}[theorem]{Lemma}
\newtheorem{proposition}[theorem]{Proposition}

\theoremstyle{definition}
\newtheorem{definition}[theorem]{Definition}
\theoremstyle{plain}

\algrenewcommand\algorithmicrequire{\textbf{Input:}}
\algrenewcommand\algorithmicensure{\textbf{Output:}}

\DeclarePairedDelimiter\rbra{\lparen}{\rparen}
\DeclarePairedDelimiter\sbra{\lbrack}{\rbrack}
\DeclarePairedDelimiter\cbra{\{}{\}}
\DeclarePairedDelimiter\abs{\lvert}{\rvert}
\DeclarePairedDelimiter\Abs{\lVert}{\rVert}

\DeclarePairedDelimiter\ket{\lvert}{\rangle}
\DeclarePairedDelimiter\bra{\langle}{\rvert}

\newcommand{\tr} {\operatorname{Tr}}

\newcommand{\spec} {\operatorname{spec}}

\DeclareMathOperator{\dist}{dist}

\newcommand{\R}{\mathbb R}
\newcommand{\C}{\mathbb C}

\newcommand{\Z}{\mathbb Z}

\newcommand{\eps}{\varepsilon}

\newcommand{\calE}{\mathcal{E}}

\newcommand{\calH}{\mathcal{H}}

\newcommand{\calL}{\mathcal{L}}

\newcommand{\cU}{\mathcal{U}}
\newcommand{\cV}{\mathcal{V}}
\newcommand{\T}{\mathbb{T}}

\newcommand{\ketbra}[2]{|#1\rangle\!\langle#2|}

\title{Optimal Diamond Distance Estimation for Unitary Channels}
\author{Jinge Bao\thanks{University of Edinburgh. \href{mailto:jinge.bao@ed.ac.uk}{\nolinkurl{jinge.bao@ed.ac.uk}}} \and Qisheng Wang\thanks{Shanghai Jiao Tong University. \href{mailto:QishengWang1994@gmail.com}{\nolinkurl{QishengWang1994@gmail.com}}}}
\date{}
\begin{document}
\maketitle

\begin{abstract}
The diamond distance between two quantum channels determines the optimal probability of distinguishing them from a single use.
We give a quantum algorithm that estimates the diamond distance between two unknown $d$-dimensional unitary channels to within additive error $\varepsilon$, using $O(\sqrt{d}/\varepsilon)$ queries to the unitary channels and their inverses.
The query complexity is \textit{optimal}, as it matches the lower bound due to \hyperlink{cite.JeonOh2026}{Jeon and Oh (\textit{npj Quantum Inf.}\ 2026)}.
\end{abstract}

\section{Introduction}

The diamond distance between quantum channels, also known as the completely bounded trace norm distance \cite{Watrous2018},
was introduced in \cite{AharonovKitaevNisan1998} to analyze quantum circuits with mixed states.
It determines the optimal probability of distinguishing two channels from a single use (cf.\ \cite{Watrous2018}).
The implementation precision of Hamiltonian simulation algorithms is often stated in terms of the diamond distance (cf.\ \cite{KLL+17,Cam19,BCS+20}).

We study the query complexity of estimating, to within additive error $\eps$, the diamond distance between the unitary channels $\cU$ and $\cV$ induced by unknown $d\times d$ unitary operators $U$ and $V$,
given query access to $U$, $V$, and their inverses.
Known classical algorithms for computing the diamond norm are based on semidefinite programming \cite{Wat09} and convex programming \cite{BATS10},
where they take explicit matrix descriptions of the channels as input.
Channel tomography in diamond distance, e.g., \cite{YRC20,HaahKothariODonnellTang2023,CGO+26}, provides such descriptions,
but learning a unitary channel this way requires $\Theta\rbra{d^2/\eps}$ queries \cite{HaahKothariODonnellTang2023}.
The spectrum estimator of \cite{HaahKothariODonnellTang2023} approximates the eigenphases of $U^\dag V$ up to a common shift using $O\rbra{d\log^2 d/\eps}$ queries,
which gives a diamond distance estimator because the diamond distance is invariant under such a shift \cite{Acin2001}.
On the other hand, the results of \cite{JeonOh2026} imply that $\Omega\rbra{\sqrt{d}/\eps}$ queries are necessary.
This leaves a gap between $O\rbra{d\log^2 d/\eps}$ and $\Omega\rbra{\sqrt{d}/\eps}$.

\subsection{Our result}

Our main result is a query-optimal estimator, which closes this gap.

\begin{theorem} \label{thm:main}
    Let $d\ge2$ and $\eps\in\interval[open left]{0}{1/4}$.
    Let $U$ and $V$ be unknown $d\times d$ unitary operators with unitary channels $\cU\rbra{X}=UXU^\dag$ and $\cV\rbra{X}=VXV^\dag$.
    The diamond distance $\Abs{\cU-\cV}_\diamond$ can be estimated to within additive error $\eps$ with probability at least $2/3$ using $O\rbra{\sqrt{d}/\eps}$ queries to $U$, $V$, $U^\dag$, and $V^\dag$.
\end{theorem}

\Cref{thm:main} immediately implies an approach to unitary channel certification considered in \cite{FFGO23,JeonOh2026,ChenWangZhang2026}, which asks to decide whether $\cU=\cV$ or $\Abs{\cU-\cV}_\diamond \geq \varepsilon$, promised that one of them holds, using $O\rbra{\sqrt{d}/\eps}$ queries to $U$, $V$, $U^\dag$, and $V^\dag$, thereby reproducing the upper bound in \cite{JeonOh2026} by a different method.
Moreover, \Cref{thm:main} is optimal up to a constant factor since it matches the lower bound for unitary channel certification given in \cite{JeonOh2026}.

\begin{figure}[!htbp]
  \centering
  % Definition of the spectral width capped at pi. Requires TikZ and amsmath.
% The manuscript provides \rbra.
% Left: phases -45,-15,25,65 degrees, shortest containing arc 110 degrees.
% Right: phases 20,100,220 degrees, shortest containing arc 200 degrees.
\begingroup
\input{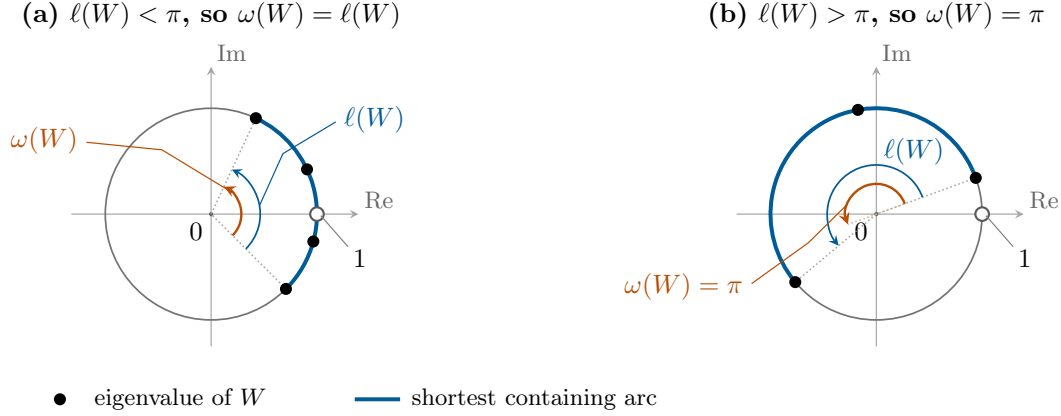}
\begin{tikzpicture}[geometry figure]
  \path[use as bounding box] (0,0.55) rectangle (\figWidth,\figTopY);
  \node[panel title] at (\figLeftX,\figTitleY) {(a) $\ell\rbra{W}<\pi$, so $\omega\rbra{W}=\ell\rbra{W}$};
  \node[panel title] at (\figRightX,\figTitleY) {(b) $\ell\rbra{W}>\pi$, so $\omega\rbra{W}=\pi$};

  \foreach \centerx in {\figLeftX,\figRightX} {
    \begin{scope}[shift={(\centerx,\figCenterY)}]
      \figureAxes
      \fill[origin fill] (0,0) circle[radius=\figOriginRadius];
      \node[below left,inner sep=3pt] at (0,0) {$0$};
    \end{scope}
  }

  \begin{scope}[shift={(\figLeftX,\figCenterY)}]
    \draw[radius guide] (-45:\figRadius) -- (0,0) -- (65:\figRadius);
    \draw[containing arc] (-45:\figRadius)
      arc[start angle=-45,end angle=65,radius=\figRadius];
    \draw[angle arc] (-45:\figAngleRadius)
      arc[start angle=-45,end angle=65,radius=\figAngleRadius];
    % Equal angles on distinct radii keep omega and ell visually separate.
    \draw[capped angle] (-45:\figCapRadius)
      arc[start angle=-45,end angle=65,radius=\figCapRadius];
    \draw[blue label guide]
      (10:\figAngleRadius) -- (0.95,0.80) -- (1.65,1.22);
    \node[anchor=west,text=figureblue,inner sep=1pt]
      at (1.72,1.25) {$\ell\rbra{W}$};
    \draw[orange label guide]
      (45:\figCapRadius) -- (-0.65,0.85) -- (-1.65,0.95);
    \node[anchor=east,text=figureorange,inner sep=1pt]
      at (-1.72,0.98) {$\omega\rbra{W}$};
    \foreach \angle in {-45,-15,25,65} {
      \node[spectral point] at (\angle:\figRadius) {};
    }
    \node[cut point] at (\figRadius,0) {};
    \draw[label guide] (1.45,-0.02) -- (1.81,-0.43);
    \node[anchor=north west,inner sep=0pt] at (1.87,-0.46) {$1$};
  \end{scope}

  \begin{scope}[shift={(\figRightX,\figCenterY)}]
    \draw[radius guide] (20:\figRadius) -- (0,0) -- (220:\figRadius);
    \draw[containing arc] (20:\figRadius)
      arc[start angle=20,end angle=220,radius=\figRadius];
    \draw[angle arc] (20:\figAngleRadius)
      arc[start angle=20,end angle=220,radius=\figAngleRadius];
    % The capped angle shares the starting radius but ends at 200 degrees.
    % This semicircle represents an angle, not a containing arc of the spectrum.
    \draw[diameter guide]
      (20:\figCapRadius) -- (200:\figCapRadius);
    \draw[capped angle] (20:\figCapRadius)
      arc[start angle=20,end angle=200,radius=\figCapRadius];
    \draw[orange label guide]
      (160:\figCapRadius) -- (-0.90,-0.38) -- (-1.65,-0.92);
    \node[anchor=east,text=figureorange,inner sep=1pt]
      at (-1.72,-0.95) {$\omega\rbra{W}=\pi$};
    \node[text=figureblue,inner sep=1pt]
      at (0.50,0.83) {$\ell\rbra{W}$};
    \foreach \angle in {20,100,220} {
      \node[spectral point] at (\angle:\figRadius) {};
    }
    \node[cut point] at (\figRadius,0) {};
    \draw[label guide] (1.45,-0.02) -- (1.81,-0.43);
    \node[anchor=north west,inner sep=0pt] at (1.87,-0.46) {$1$};
  \end{scope}

  \legendMarker{spectral point}{\legendColA}{\legendRowA}{eigenvalue of $W$}
  \legendLine{containing arc}{\legendColB}{\legendRowA}{shortest containing arc}
\end{tikzpicture}
\endgroup
  \caption{Panels (a) and (b) show the cases $\ell\rbra{W}<\pi$ and $\ell\rbra{W}>\pi$.
  The blue angle is the length $\ell\rbra{W}$ of a shortest containing arc, and the orange angle is $\omega\rbra{W}=\min\cbra{\ell\rbra{W},\pi}$.}
  \label{fig:spectral-width-definition}
\end{figure}

The diamond distance satisfies $\Abs{\cU-\cV}_\diamond=2\sin\rbra{\omega\rbra{U^\dag V}/2}$ \cite{Acin2001,HaahKothariODonnellTang2023}, where for a unitary $W$, $\omega\rbra{W}=\min\cbra{\ell\rbra{W},\pi}$ and $\ell\rbra{W}$ is the length of a shortest closed arc of the unit circle containing the spectrum of $W$, as illustrated in \cref{fig:spectral-width-definition}.
It therefore suffices to estimate $\omega\rbra{W}$ for $W=U^\dag V$.

\subsection{Related work}

Using local operations and classical communication, the algorithm of \cite{ZhengWangYuXuZhang2025} estimates the overlap between the normalized Choi states of two unknown unitary channels to additive error $\eps$
using $O\rbra{\max\cbra{\sqrt d/\eps,1/\eps^2}}$ queries.

In a maximally entangled state, the vectors of every orthonormal basis appear with equal amplitudes, and spectral estimation algorithms \cite{vanApeldoornGilyenGriblingDeWolf2020,MandeDeWolf2026,JefferyWitteveen2026} use it as input.
With this input, the algorithm of \cite{JefferyWitteveen2026} estimates the maximum eigenphase of a unitary to additive error $\eps$ using $O\rbra{\sqrt d/\eps}$ queries.
The algorithm assumes controlled access and a spectrum contained in a fixed quarter of the unit circle.
Our interval search needs neither assumption.

Like our algorithm, the spectrum estimator of \cite{HaahKothariODonnellTang2023} retains an unknown eigenvector as a reference for eigenphase differences.
Our search is the bounded-error quantum search of \cite{HoyerMoscaDeWolf2003} in the form developed in \cite{KociumakaNoglerWellnitz2025}.
Our interval search builds on \cite{LinTong2020,DongLinTong2022} and accepts phase estimates near the point where phases wrap around.

\section{Preliminaries}
\label{sec:preliminaries}

\subsection{Notation}
\label{subsec:notation}
All Hilbert spaces are finite-dimensional over $\C$.
The notation $\calL\rbra{\calH}$ denotes the space of linear operators on a Hilbert space $\calH$,
$\mathbb{U}\rbra{d}$ the group of unitaries on $\C^d$, and $I$ the identity on the space determined by context.
For a positive integer $n$, let $\sbra{n}=\cbra{0,\ldots,n-1}$.
For an operator $A$, $A^\dag$ denotes its adjoint, $\spec\rbra{A}$ its spectrum,
and $\Abs{A}_1=\tr\rbra{\sqrt{A^\dag A}}$ its trace norm.
For a vector $x$ in a Hilbert space, $\Abs{x}$ denotes its Euclidean norm.

Eigenphases of unitaries are defined modulo $2\pi$ and belong to $\T=\R/\rbra{2\pi\Z}$.
For real angles $\alpha,\beta$, their circular distance is
\begin{align*}
  \dist_{\T}\rbra*{\alpha, \beta} \coloneqq \min_{k \in \Z} \abs*{\alpha - \beta + 2\pi k}.
\end{align*}
Unless otherwise stated, we represent each eigenvalue as $e^{i\phi}$ using the unique angle $\phi\in\interval[open right]{0}{2\pi}$.
All maxima, minima, and comparisons of eigenphases refer to the chosen real angles.

\begin{figure}[!htbp]
  \centering
  % Vector illustration for the phase-cut discussion. Requires TikZ, amsmath, and amssymb.
% The manuscript provides \dist, \T, \rbra, and \abs.
\begingroup
\input{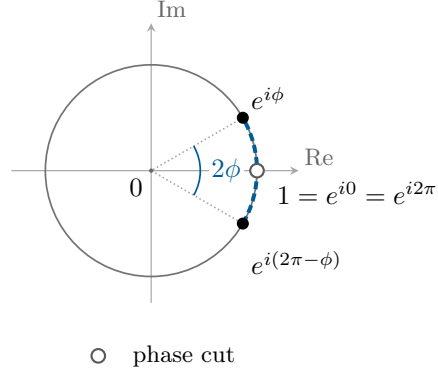}
\begin{tikzpicture}[geometry figure,short arc/.append style={dash pattern=on 3pt off 1.6pt}]
  \path[use as bounding box] (0,0.55) rectangle (\figWidth,5.40);

  \begin{scope}[shift={(\figMiddleX,\figCenterY)}]
    \coordinate (upper) at (30:\figRadius);
    \coordinate (lower) at (-30:\figRadius);
    \figureAxes
    \draw[radius guide] (upper) -- (0,0) -- (lower);
    \draw[short arc] (-30:\figRadius) arc[start angle=-30,end angle=30,radius=\figRadius];
    \draw[angle measure] (-30:\figAngleRadius) arc[start angle=-30,end angle=30,radius=\figAngleRadius];
    \node[text=figureblue,fill=white,inner sep=1pt] at (0.99,0) {$2\phi$};
    \node[point] at (upper) {};
    \node[point] at (lower) {};
    \node[above right,inner sep=3pt] at (upper) {$e^{i\phi}$};
    \node[below right,inner sep=3pt,yshift=-7pt] at (lower) {$e^{i(2\pi-\phi)}$};
    \node[cut point] at (\figRadius,0) {};
    \node[anchor=north west,align=left,inner sep=0pt] at (1.66,-0.14)
      {$1=e^{i0}=e^{i2\pi}$};
    \fill[origin fill] (0,0) circle[radius=\figOriginRadius];
    \node[below left,inner sep=3pt] at (0,0) {$0$};
  \end{scope}

  % A single entry is centered under the circle.
  \legendMarker{cut point}{\figMiddleX-1.0}{\legendRowA}{phase cut}
\end{tikzpicture}
\endgroup
  \caption{Phase representatives in $\interval[open right]{0}{2\pi}$ wrap around at the phase cut, the point $1$ on the unit circle.
  The phases $\phi$ and $2\pi-\phi$ with $0<\phi<\pi/2$ have circular distance $2\phi$ but absolute difference $2\pi-2\phi$.}
  \label{fig:phase-cut-distances}
\end{figure}

The chosen angle jumps from near $2\pi$ to near $0$ as the eigenvalue crosses $1$ counterclockwise on the unit circle.
We call the phase $0\in\T$, corresponding to the point $1$ on the unit circle, the \emph{phase cut}.
\Cref{fig:phase-cut-distances} illustrates circular distance across it.

Unless otherwise stated, logarithms are natural.
Register names are typeset in uppercase sans serif, and a subscript, as in $\ket{\psi}_{\mathsf{A}}$, indicates the register that holds a state.

\subsection{Diamond distance}
\begin{definition}[{Diamond norm, \cite{Watrous2018}}] \label{def:diamond-norm}
    Let $\mathsf{S}$ and $\mathsf{R}$ be $d$-dimensional Hilbert spaces, and let $\mathrm{id}_{\mathsf{R}}$ be the identity channel on $\mathsf{R}$.
    For a linear map $\calE \colon \calL\rbra{\mathsf{S}} \to \calL\rbra{\mathsf{S}}$, its diamond norm is
    \begin{align*}
        \Abs*{\calE}_\diamond = \max_{\substack{A \in \calL\rbra*{\mathsf{S} \otimes \mathsf{R}} \\ \Abs*{A}_1=1}}\Abs*{\rbra*{\calE \otimes \mathrm{id}_{\mathsf{R}}} \rbra*{A} }_1.
    \end{align*}
\end{definition}
Taking $\mathsf R$ of dimension larger than $d$ does not increase the maximum~\cite[Theorem~3.46]{Watrous2018}.
\begin{definition}[Diamond distance] \label{def:diamond distance}
    For linear maps $\calE_1,\calE_2\colon\calL\rbra{\mathsf{S}}\to\calL\rbra{\mathsf{S}}$, their diamond distance is $\Abs{\calE_1-\calE_2}_\diamond$.
\end{definition}
For $U,V\in\mathbb{U}\rbra{d}$, let $\cU\rbra{X}=UXU^\dag$ and $\cV\rbra{X}=VXV^\dag$ be the induced unitary channels.
For $W\in\mathbb{U}\rbra{d}$, its spectral width $\ell\rbra{W}\in\interval{0}{2\pi}$
is the length of a shortest closed arc containing $\spec\rbra{W}$,
and its truncated spectral width is $\omega\rbra{W}=\min\cbra{\ell\rbra{W},\pi}$.
\begin{proposition}[{Spectral-width formula, \cite{Acin2001,HaahKothariODonnellTang2023}}] \label{prop:width-formula}
    For $U,V\in \mathbb{U}\rbra{d}$,
    \begin{align*}
        \Abs*{\cU-\cV}_\diamond = 2\sin\rbra*{\frac{\omega\rbra*{U^\dag V}}{2}}.
    \end{align*}
\end{proposition}
Equivalent characterizations appear in \cite{AharonovKitaevNisan1998,JW09,KR21}.

\subsection{Computational model}
\label{subsec:query-model}
A \emph{query} is one application of $U$, $V$, $U^\dag$, or $V^\dag$.
A $q$-query quantum algorithm has the form
\begin{align*}
    A_q O_q A_{q-1}O_{q-1}\cdots A_1O_1A_0.
\end{align*}
Each $O_i$ applies one of these four unitaries to a designated register, and each $A_i$ is a unitary independent of the inputs.
More generally, an algorithm may measure registers during the computation and choose its subsequent operations based on the outcomes and on classical randomness.
The query complexity of an algorithm is the maximum number of queries over all inputs and all execution branches.
Let $W=U^\dag V$ be the relative unitary. Each application of $W$ or $W^\dag$ uses two queries.

\subsection{Quantum subroutines} \label{subsec:quantum-subroutines}
For an orthonormal basis $\ket{v_0},\ldots,\ket{v_{n-1}}$ of $\C^n$, let
$\ket{\chi_j}=\ket{v_j}\ket{v_j^\ast}$, where $\ket{v_j^\ast}$ denotes the complex conjugate of $\ket{v_j}$ in the computational basis.
Their uniform superposition is the maximally entangled state
\begin{equation}
  \ket{\Phi_n}
  =\frac1{\sqrt n}\sum_{k=0}^{n-1}\ket{k}\ket{k}
  =\frac1{\sqrt n}\sum_{j=0}^{n-1}\ket{v_j}\ket{v_j^\ast}
  =\frac1{\sqrt n}\sum_{j=0}^{n-1}\ket{\chi_j}.
  \label{eq:epr-decomposition}
\end{equation}
By the first expression in \cref{eq:epr-decomposition}, input-independent circuits prepare $\ket{\Phi_n}$ and reflect about it.

\begin{definition}[Label-preserving verifier] \label{def:label-preserving}
  Let $\ket{v_0},\ldots,\ket{v_{n-1}}$ be an orthonormal basis of $\C^n$, and let $\ket{\chi_j}=\ket{v_j}\ket{v_j^\ast}$.
  A \emph{label-preserving verifier} on these labels is a unitary circuit $F$ with a flag qubit $\mathsf Q$ in its workspace such that
  \begin{align*}
    F\rbra*{\ket{\chi_j}\otimes\ket{w}}=\ket{\chi_j}\otimes F_j\ket{w}
  \end{align*}
  for every $j\in\sbra{n}$ and workspace state $\ket{w}$, where each $F_j$ is unitary.
  Its \emph{acceptance probability} on label $j$ is
  \begin{align*}
    p_j\coloneqq\Abs*{\rbra{I\otimes\bra{1}_{\mathsf Q}}F_j\ket{0}}^2,
  \end{align*}
  the probability of measuring $1$ in $\mathsf Q$ when the workspace starts in $\ket{0}$, where $I$ acts on the remaining workspace.
\end{definition}

\begin{lemma}[{Adapted from \cite[Lemma~3.5]{KociumakaNoglerWellnitz2025}}]
  \label{lem:neutral-search}
  Let $F$ be a label-preserving verifier on $n$ labels.
  For every $\delta\in\interval[open]{0}{1/3}$, there is an algorithm $\textnormal{\textsc{Search}}\rbra{F,\delta}$ that uses input-independent gates and
  $O\rbra{\sqrt n\log\rbra{1/\delta}}$ applications of $F$ and $F^\dag$ and outputs $b\in\cbra{0,1}$.
  With probability at least $1-\delta$, $b=1$ if $\max_{j\in\sbra{n}}p_j\ge9/10$, and $b=0$ if $\max_{j\in\sbra{n}}p_j<1/10$.
\end{lemma}

\begin{proof}
We run the constant-error construction of \cite[Lemma~3.5]{KociumakaNoglerWellnitz2025}
with thresholds $9/10$ and $1/10$, and record only whether it reports an index.
That construction acts on computational-basis indices $\ket{i}$, $i\in\sbra{n}$, and uses the index register only through the oracle,
through amplitude amplification from the uniform superposition of indices,
and through measurements of the index followed by repeated oracle calls on the measured index.
Let $T$ be the isometry with $T\ket{i}=\ket{\chi_i}$.
By \cref{eq:epr-decomposition}, $T$ maps the uniform superposition of indices to $\ket{\Phi_n}$,
and by label preservation, $F\rbra{T\otimes I}=\rbra{T\otimes I}\sum_{i}\ketbra{i}{i}\otimes F_i$.
Hence $F$ acts on the labels as an index oracle with acceptance probabilities $p_i$.
On the span of the label states, the reflections used by amplitude amplification act in the same way as the corresponding reflections on indices under $T\otimes I$.
Two changes handle the remaining operations.

First, the error-reduction steps of that construction call the oracle only on branches whose current flag is $1$.
We instead apply each such verification to all branches on fresh workspace and store the AND of its result and the current flag in a fresh qubit.
For every label, the component whose new flag is $1$ coincides with that of the original construction, up to ancilla registers in fixed states.
With all workspace retained, amplitude amplification of the full modified circuit and its inverse rescales the flag-$1$ component by a factor that depends only on its norm.
By induction over the recursion, the analysis of \cite{KociumakaNoglerWellnitz2025} therefore applies with applications of $F$ and $F^\dag$.
Second, instead of measuring the label, we apply the repeated verifications to the unmeasured label register and measure only their flags.
These verifications are block diagonal with respect to the labels,
so the flags have the same joint distribution as when the label is measured first.

The resulting decision succeeds with probability at least $2/3$ under either promise using $O\rbra{\sqrt n}$ applications of $F$ and $F^\dag$.
A majority vote over $O\rbra{\log\rbra{1/\delta}}$ independent repetitions reduces the failure probability to at most $\delta$.
\end{proof}

\begin{lemma}[{Quantum phase estimation, \cite{Kitaev1995,MandeDeWolf2026}}]
  \label{lem:phase-estimation}
  Let $Y\in\mathbb U\rbra{N}$, $s\in\interval[open left]{0}{1}$, and $\delta\in\interval[open]{0}{1/3}$.
  There is a unitary circuit that
  uses $O\rbra{\log\rbra{1/\delta}/s}$ controlled applications of $Y$ or $Y^\dag$.
  The circuit leaves each input eigenvector of $Y$ unchanged for every initial workspace state.
  On input $\ket{\psi}\ket{0}$ with $Y\ket{\psi}=e^{i\phi}\ket{\psi}$,
  measuring its output phase register yields, with probability at least $1-\delta$,
  an estimate $\widetilde\phi\in\interval[open right]{0}{2\pi}$ with $\dist_{\T}\rbra{\widetilde\phi,\phi}\le s$.
\end{lemma}

\section{Proof of the main theorem}
\label{sec:upper-bound}

We prove \cref{thm:main} by estimating the truncated spectral width in \cref{prop:width-formula}.
Throughout, $U,V\in\mathbb{U}\rbra{d}$ are the input unitaries, with $d\ge2$.

\subsection{Shifted relative eigenphases} \label{subsec:relative-eigenphases}

Let the relative unitary $W$ have an orthonormal eigenbasis $\ket{u_j}$ with
$W\ket{u_j}=e^{i\theta_j}\ket{u_j}$, where
$\theta_j\in\interval[open right]{0}{2\pi}$ and $j\in\sbra{d}$.
For $j,r\in\sbra{d}$, define the \emph{shifted relative eigenphases}
\begin{align*}
  \phi_{j,r}\coloneqq\rbra{\theta_j-\theta_r+\pi}\bmod 2\pi
  \in\interval[open right]{0}{2\pi}.
\end{align*}
We call $r$ the reference index. The shift places $\phi_{r,r}$ at $\pi$ and sends the antipode of $e^{i\theta_r}$ to the phase cut.

\begin{lemma}[Truncated spectral width]
  \label{lem:relative-width}
  For every $r\in\sbra{d}$,
  \begin{align*}
    \omega\rbra*{W}=\min\cbra*{ \max_{j\in\sbra*{d}}\phi_{j,r}-\min_{j\in\sbra*{d}}\phi_{j,r},\pi}.
  \end{align*}
\end{lemma}

\begin{proof}
Fix $r\in\sbra{d}$.\nopagebreak

\emph{Case 1: $\ell\rbra{W}<\pi$.}
See \cref{fig:relative-phase-geometry}.
\begin{figure}[!ht]
  \centering
  % Vector illustration for Section 4.1. Requires TikZ and amsmath.
% The manuscript provides \rbra.
% The example has phases -35,-10,20,50 degrees; the reference is 20 degrees.
% Rotating coordinates by 160 degrees gives 125,150,180,210 degrees.
% The rotation is an analytical change of coordinates, not an oracle operation.
\begingroup
\input{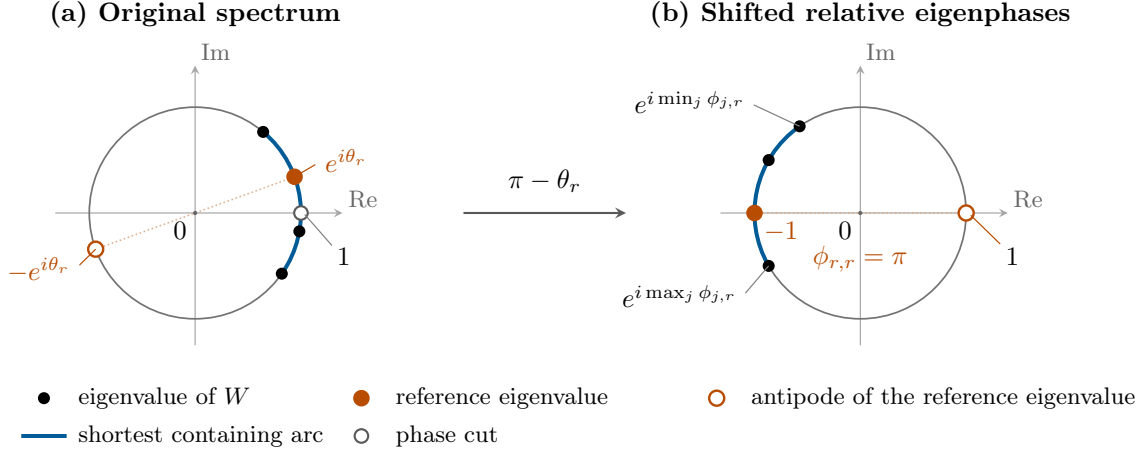}
\begin{tikzpicture}[geometry figure]
  \path[use as bounding box] (0,0.05) rectangle (\figWidth,\figTopY);
  \node[panel title] at (\figLeftX,\figTitleY) {(a) Original spectrum};
  \node[panel title] at (\figRightX,\figTitleY) {(b) Shifted relative eigenphases};

  \begin{scope}[shift={(\figLeftX,\figCenterY)}]
    \figureAxes
    \draw[diameter guide] (200:\figRadius) -- (20:\figRadius);
    \draw[short arc] (-35:\figRadius) arc[start angle=-35,end angle=50,radius=\figRadius];
    \foreach \angle in {-35,-10,50} {
      \node[point] at (\angle:\figRadius) {};
    }
    \node[reference point] at (20:\figRadius) {};
    \node[antipode point] at (200:\figRadius) {};
    \node[anchor=west,text=figureorange,inner sep=1pt]
      at (1.68,0.69) {$e^{i\theta_r}$};
    \draw[orange label guide] (20:\figRadius) -- (1.59,0.64);
    \node[anchor=east,text=figureorange,inner sep=1pt]
      at (-1.61,-0.75) {$-e^{i\theta_r}$};
    \draw[orange label guide] (200:\figRadius) -- (-1.53,-0.69);
    \node[cut point] at (\figRadius,0) {};
    \draw[label guide] (1.45,-0.02) -- (1.81,-0.43);
    \node[anchor=north west,align=left,inner sep=0pt] at (1.87,-0.46)
      {$1$};
    \fill[origin fill] (0,0) circle[radius=\figOriginRadius];
    \node[below left,inner sep=3pt] at (0,0) {$0$};
  \end{scope}

  \draw[rotation arrow] (6.85,\figCenterY) -- (9.00,\figCenterY);
  \node at (7.925,{\figCenterY+0.43}) {$\pi-\theta_r$};

  \begin{scope}[shift={(\figRightX,\figCenterY)}]
    \figureAxes
    \draw[diameter guide] (-\figRadius,0) -- (\figRadius,0);
    \draw[short arc] (125:\figRadius) arc[start angle=125,end angle=210,radius=\figRadius];
    \foreach \angle in {125,150,210} {
      \node[point] at (\angle:\figRadius) {};
    }
    \node[reference point] at (180:\figRadius) {};
    \node[antipode point] at (0:\figRadius) {};
    \draw[label guide] (125:\figRadius) -- (-1.41,1.43);
    \node[anchor=east,inner sep=1pt] at (-1.48,1.47) {$e^{i\min_j\phi_{j,r}}$};
    \draw[label guide] (210:\figRadius) -- (-1.52,-1.06);
    \node[anchor=east,inner sep=1pt] at (-1.59,-1.10) {$e^{i\max_j\phi_{j,r}}$};
    \node[anchor=north west,text=figureorange,inner sep=2pt]
      at (-1.35,-0.04) {$-1$};
    \node[text=figureorange,inner sep=1pt] at (0,-0.60) {$\phi_{r,r}=\pi$};
    \draw[orange label guide] (1.46,-0.03) -- (1.85,-0.41);
    \node[anchor=north west,align=left,inner sep=0pt] at (1.92,-0.45)
      {$1$};
    \fill[origin fill] (0,0) circle[radius=\figOriginRadius];
    \node[below left,inner sep=3pt] at (0,0) {$0$};
  \end{scope}

  \legendMarker{point}{\legendColA}{\legendRowA}{eigenvalue of $W$}
  \legendMarker{reference point}{\legendColB}{\legendRowA}{reference eigenvalue}
  \legendMarker{antipode point}{\legendColC}{\legendRowA}{antipode of the reference eigenvalue}
  \legendLine{short arc}{\legendColA}{\legendRowB}{shortest containing arc}
  \legendMarker{cut point}{\legendColB}{\legendRowB}{phase cut}
\end{tikzpicture}
\endgroup
  \caption{A shortest arc containing the spectrum of $W$, before (a) and after (b) rotation by $\pi-\theta_r$,
  which sends the reference eigenvalue $e^{i\theta_r}$ to $-1$ and its antipode to the phase cut at $1$.
  Since the arc has length $\ell\rbra{W}<\pi$, after the rotation it avoids the phase cut,
  so the angles $\phi_{j,r}\in\interval[open right]{0}{2\pi}$ of the rotated eigenvalues satisfy $\max_j\phi_{j,r}-\min_j\phi_{j,r}=\ell\rbra{W}$.}
  \label{fig:relative-phase-geometry}
\end{figure}
Rotating a shortest containing arc by $\pi-\theta_r$ gives an arc of length $\ell\rbra{W}<\pi$
containing $-1$, the image of $e^{i\theta_r}$, so it avoids the cut at $1$.
The angles in $\interval[open right]{0}{2\pi}$ corresponding to points of this arc form a real interval of length $\ell\rbra{W}$.
The shifted phases attain both endpoints, since otherwise the containing arc could be shortened.
Hence $\omega\rbra{W} = \ell\rbra{W} = \max_j\phi_{j,r}-\min_j\phi_{j,r}$.

\emph{Case 2: $\ell\rbra{W}\ge\pi$.}
See \cref{fig:relative-phase-geometry-wide}.
\begin{figure}[!ht]
  \centering
  % Vector illustration for the regime ell(W) >= pi. Requires TikZ and amsmath.
% The manuscript provides \rbra.
% Original eigenphases: 20,100,220 degrees; reference phase: 20 degrees.
% A coordinate shift by 160 degrees gives canonical phases 180,260,20 degrees.
% The shortest containing arc has length 200 degrees in both coordinate systems.
% The phase interval has width 240 degrees, strictly larger here.
\begingroup
\input{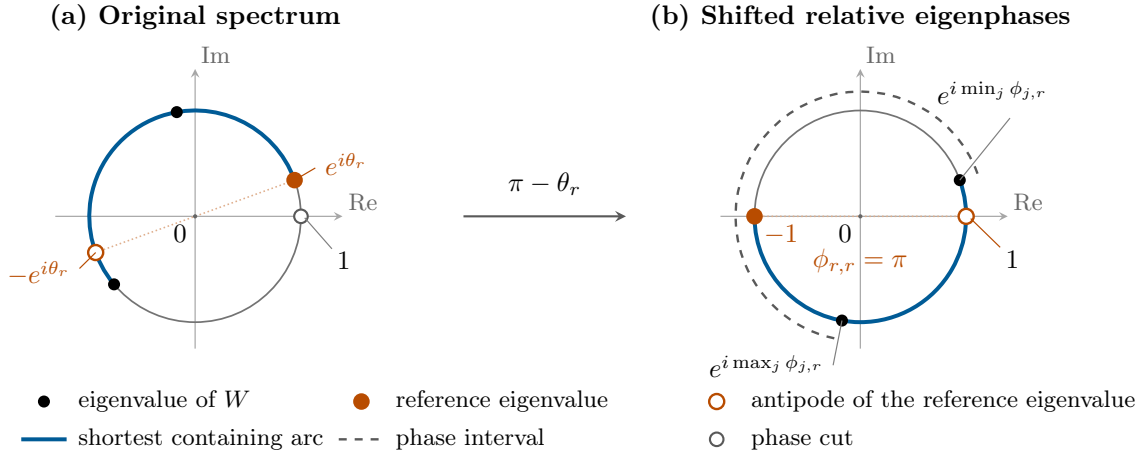}
\begin{tikzpicture}[geometry figure]
  \path[use as bounding box] (0,0.05) rectangle (\figWidth,\figTopY);
  \node[panel title] at (\figLeftX,\figTitleY) {(a) Original spectrum};
  \node[panel title] at (\figRightX,\figTitleY) {(b) Shifted relative eigenphases};

  \begin{scope}[shift={(\figLeftX,\figCenterY)}]
    \figureAxes
    \draw[diameter guide] (200:\figRadius) -- (20:\figRadius);
    \draw[short arc] (20:\figRadius) arc[start angle=20,end angle=220,radius=\figRadius];
    \foreach \angle in {100,220} {
      \node[point] at (\angle:\figRadius) {};
    }
    \node[reference point] at (20:\figRadius) {};
    \node[antipode point] at (200:\figRadius) {};
    \node[anchor=west,text=figureorange,inner sep=1pt]
      at (1.68,0.69) {$e^{i\theta_r}$};
    \draw[orange label guide] (20:\figRadius) -- (1.59,0.64);
    \node[anchor=east,text=figureorange,inner sep=1pt]
      at (-1.61,-0.75) {$-e^{i\theta_r}$};
    \draw[orange label guide] (200:\figRadius) -- (-1.53,-0.69);
    \node[cut point] at (\figRadius,0) {};
    \draw[label guide] (1.45,-0.02) -- (1.81,-0.43);
    \node[anchor=north west,align=left,inner sep=0pt] at (1.87,-0.46)
      {$1$};
    \fill[origin fill] (0,0) circle[radius=\figOriginRadius];
    \node[below left,inner sep=3pt] at (0,0) {$0$};
  \end{scope}

  \draw[rotation arrow] (6.85,\figCenterY) -- (9.00,\figCenterY);
  \node at (7.925,{\figCenterY+0.43}) {$\pi-\theta_r$};

  \begin{scope}[shift={(\figRightX,\figCenterY)}]
    \figureAxes
    \draw[diameter guide] (-\figRadius,0) -- (\figRadius,0);
    \draw[short arc] (180:\figRadius) arc[start angle=180,end angle=380,radius=\figRadius];
    \draw[phase interval] (20:\figIntervalRadius) arc[start angle=20,end angle=260,radius=\figIntervalRadius];
    \foreach \angle in {20,260} {
      \node[point] at (\angle:\figRadius) {};
    }
    \node[reference point] at (180:\figRadius) {};
    \node[antipode point] at (0:\figRadius) {};
    \draw[label guide] (20:\figRadius) -- (1.98,1.30);
    \node[anchor=south,inner sep=1pt] at (1.75,1.45)
      {$e^{i\min_j\phi_{j,r}}$};
    \draw[label guide] (260:\figRadius) -- (-0.36,-1.97);
    \node[anchor=east,inner sep=1pt] at (-0.44,-1.98)
      {$e^{i\max_j\phi_{j,r}}$};
    \node[anchor=north west,text=figureorange,inner sep=2pt]
      at (-1.35,-0.04) {$-1$};
    \node[text=figureorange,inner sep=1pt] at (0,-0.60) {$\phi_{r,r}=\pi$};
    \draw[orange label guide] (1.46,-0.03) -- (1.85,-0.41);
    \node[anchor=north west,align=left,inner sep=0pt] at (1.92,-0.45)
      {$1$};
    \fill[origin fill] (0,0) circle[radius=\figOriginRadius];
    \node[below left,inner sep=3pt] at (0,0) {$0$};
  \end{scope}

  \legendMarker{point}{\legendColA}{\legendRowA}{eigenvalue of $W$}
  \legendMarker{reference point}{\legendColB}{\legendRowA}{reference eigenvalue}
  \legendMarker{antipode point}{\legendColC}{\legendRowA}{antipode of the reference eigenvalue}
  \legendLine{short arc}{\legendColA}{\legendRowB}{shortest containing arc}
  \legendLine{phase interval}{\legendColB}{\legendRowB}{phase interval}
  \legendMarker{cut point}{\legendColC}{\legendRowB}{phase cut}
\end{tikzpicture}
\endgroup
  \caption{Rotation by $\pi-\theta_r$ when a shortest arc containing the spectrum of $W$ has length $\ell\rbra{W}\ge\pi$.
  After the rotation, this arc may contain the phase cut at $1$.
  The phase interval corresponds to $\interval{\min_j\phi_{j,r}}{\max_j\phi_{j,r}}$ and has length at least $\ell\rbra{W}$.
  In this example the two lengths are $\ell\rbra{W}=10\pi/9$ and $4\pi/3$, so both truncate to $\pi$.}
  \label{fig:relative-phase-geometry-wide}
\end{figure}
The interval $\interval{\min_j\phi_{j,r}}{\max_j\phi_{j,r}}$
parametrizes an arc of length less than $2\pi$.
Rotating it by $\theta_r-\pi$ gives an arc containing $\spec\rbra{W}$, so
\begin{align*}
  \max_j\phi_{j,r}-\min_j\phi_{j,r}\ge\ell\rbra*{W}\ge\pi.
\end{align*}
Truncation therefore gives $\pi=\omega\rbra{W}$.
\end{proof}

We next construct a \emph{routing circuit} that encodes the shifted relative eigenphases on a control qubit.
Take a reference register $\mathsf A$ and a target register $\mathsf B$, each of dimension $d$, and a control qubit $\mathsf C$.
Let $\operatorname{SWAP}_{\mathsf{AB}}$ exchange $\mathsf A$ and $\mathsf B$, and let
$\operatorname{c-SWAP}=\ketbra{0}{0}_{\mathsf C}\otimes I_{\mathsf A}\otimes I_{\mathsf B}+\ketbra{1}{1}_{\mathsf C}\otimes\operatorname{SWAP}_{\mathsf{AB}}$.
The routing circuit applies $\operatorname{c-SWAP}$, $I_{\mathsf C}\otimes W_{\mathsf A}\otimes I_{\mathsf B}$, $\operatorname{c-SWAP}$, and $Z_{\mathsf C}\otimes I_{\mathsf A}\otimes I_{\mathsf B}$ in this order, where $Z_{\mathsf C}$ is the Pauli gate.
It is the circuit in the proof of \cite[Proposition~B.3]{HaahKothariODonnellTang2023} with $Z_{\mathsf C}$ added, which adds $\pi$ to each difference $\theta_j-\theta_r$.
Since $\operatorname{SWAP}_{\mathsf{AB}}\rbra{W_{\mathsf A}\otimes I_{\mathsf B}}\operatorname{SWAP}_{\mathsf{AB}}=I_{\mathsf A}\otimes W_{\mathsf B}$, the routing circuit equals
\begin{equation}
\begin{aligned}
  &\rbra{Z_{\mathsf C}\otimes I_{\mathsf A}\otimes I_{\mathsf B}}\operatorname{c-SWAP}\rbra{I_{\mathsf C}\otimes W_{\mathsf A}\otimes I_{\mathsf B}}\operatorname{c-SWAP}\\
  &\qquad=\ketbra{0}{0}_{\mathsf C}\otimes W_{\mathsf A}\otimes I_{\mathsf B}-\ketbra{1}{1}_{\mathsf C}\otimes I_{\mathsf A}\otimes W_{\mathsf B}.
\end{aligned}
\label{eq:routing}
\end{equation}
Because $-e^{i\theta_j}=e^{i\theta_r}e^{i\phi_{j,r}}$, it maps $\ket{c}_{\mathsf C}\ket{u_r}_{\mathsf A}\ket{u_j}_{\mathsf B}$ to $e^{i\theta_r}e^{ic\phi_{j,r}}$ times itself
for all $c\in\cbra{0,1}$ and $j,r\in\sbra{d}$, and it uses two queries.

\subsection{Estimating the maximum shifted relative eigenphase}
\label{subsec:fuzzy-maximum}

Fix $\sigma\in\cbra{+1,-1}$ and a reference index $r$, and define $\phi_j\coloneqq\rbra{\sigma\phi_{j,r}}\bmod 2\pi\in\interval[open right]{0}{2\pi}$.
For $\sigma=+1$, we have $\phi_j=\phi_{j,r}$.
For $\sigma=-1$, each positive phase $\phi_{j,r}$ becomes $2\pi-\phi_{j,r}$, while zero remains zero.
We omit the dependence on $\sigma$ and $r$ in the notation $\phi_j$ and $M$.
To treat a phase at the cut as the upper endpoint $2\pi$, define
\begin{align*}
  M\coloneqq\begin{cases}
    2\pi, & \text{if }\phi_j=0\text{ for some }j,\\
    \max_j\phi_j, & \text{otherwise}.
  \end{cases}
\end{align*}
Since $\phi_r=\pi$, we have $M\in\interval{\pi}{2\pi}$.
Throughout, fix $\eps\in\interval[open left]{0}{1/4}$, and let $\gamma=\sqrt\eps/4$ and $h=\eps/4$.

For a threshold $t$, we say that an estimate $\widetilde\phi\in\interval[open right]{0}{2\pi}$ is \emph{accepted}
if $\widetilde\phi\ge t$ or $\dist_{\T}\rbra{\widetilde\phi,0}\le\gamma$.
When the reference register $\mathsf A$ is in $\ket{u_r}$, the routing circuit with $W$ replaced by $W^\sigma$ equals $e^{i\sigma\theta_r}$ times the controlled version of $Y=-e^{-i\sigma\theta_r}W^\sigma$ on $\mathsf B$,
and $Y\ket{u_j}=e^{i\phi_j}\ket{u_j}$ because $\sigma\rbra{\theta_j-\theta_r}+\pi\equiv\sigma\rbra{\theta_j-\theta_r+\pi}\pmod{2\pi}$.
For $\sigma\in\cbra{+1,-1}$, $s\in\interval[open left]{0}{1}$, and a threshold $t$, the verifier $F^\sigma_{s,t}$ runs phase estimation for $Y$ on $\mathsf B$ with precision $s$ and failure probability at most $1/20$.
It uses this routing circuit and its inverse for the calls required by phase estimation.
The inverse is obtained by replacing $W^\sigma$ with $W^{-\sigma}$.
Finally, the verifier flips the flag qubit $\mathsf Q$ if the estimate in the phase register is accepted.

Let $\mathsf B'$ be a further $d$-dimensional register, and define the target labels $\ket{\chi_j}=\ket{u_j}_{\mathsf B}\ket{u_j^\ast}_{\mathsf B'}$.
By \cref{eq:epr-decomposition}, their uniform superposition is $\ket{\Phi_d}_{\mathsf{BB'}}$.
Each search run starts with $\mathsf{BB'}$ and its workspace in the zero state,
and the preparation of $\ket{\Phi_d}$, its inverse, and the reflections act only on these registers.

\begin{lemma}[Cut-aware comparison verifier]
  \label{lem:cut-verifier}
  Let $\sigma\in\cbra{+1,-1}$, $r\in\sbra{d}$, $s\in\interval[open left]{0}{1}$, and $t\in\R$, and let $\mathsf A$ be in the state $\ket{u_r}$.
  For fixed $r$, $F^\sigma_{s,t}$ is a label-preserving verifier on the labels $\ket{\chi_j}$,
  and each of $F^\sigma_{s,t}$ and $\rbra{F^\sigma_{s,t}}^\dag$ uses $O\rbra{1/s}$ queries.
  If every estimate $\widetilde\phi$ with $\dist_{\T}\rbra{\widetilde\phi,\phi_j}\le s$ is accepted, then the acceptance probability of label $j$ satisfies $p_j\ge19/20$, and if no such $\widetilde\phi$ is accepted, then $p_j\le1/20$.
\end{lemma}

\begin{proof}
All verifier workspace remains unmeasured, and reversing the full circuit implements $\rbra{F^\sigma_{s,t}}^\dag$.
By \cref{lem:phase-estimation} and \cref{eq:routing}, $F^\sigma_{s,t}$ preserves each target eigenvector for every workspace input.
For fixed $r$, the overall phase can be included in the workspace unitaries $F_j$ of \cref{def:label-preserving}, so $F^\sigma_{s,t}$ is label-preserving.
By \cref{lem:phase-estimation}, phase estimation uses $O\rbra{1/s}$ applications of the routing circuit or its inverse, each requiring two queries.

Phase estimation maps $\ket{\chi_j}\ket{0}$ to $\ket{\chi_j}\sum_{\widetilde\phi}\ket{\widetilde\phi}\ket{\xi_{j,\widetilde\phi}}$,
where the sum ranges over the phase-register values and the subnormalized vectors $\ket{\xi_{j,\widetilde\phi}}$ contain the remaining workspace.
Let $\mathcal S_t$ be the set of accepted values and $\mathcal N_j=\cbra{\widetilde\phi:\dist_{\T}\rbra{\widetilde\phi,\phi_j}\le s}$.
Since $F^\sigma_{s,t}$ flips $\mathsf Q$ exactly on $\mathcal S_t$,
\begin{align*}
  p_j=\sum_{\widetilde\phi\in\mathcal S_t}\Abs*{\ket{\xi_{j,\widetilde\phi}}}^2,
\end{align*}
and by \cref{lem:phase-estimation},
\begin{align*}
  \sum_{\widetilde\phi\in\mathcal N_j}\Abs*{\ket{\xi_{j,\widetilde\phi}}}^2\ge19/20.
\end{align*}
If $\mathcal N_j\subseteq\mathcal S_t$, then $p_j\ge19/20$.
If $\mathcal N_j\cap\mathcal S_t=\emptyset$, then $p_j\le1/20$.
\end{proof}

\begin{lemma}[Maximum shifted relative eigenphase]
  \label{lem:fuzzy-max}
  Let $\sigma\in\cbra{+1,-1}$ and $\delta\in\interval[open]{0}{1/3}$.
  There is an algorithm that outputs an estimate $\widehat M\in\interval{\pi}{2\pi}$ using $O\rbra{\sqrt d\log\rbra{1/\delta}/\eps}$ queries in the worst case.
  For every reference index $r$, if $\mathsf A$ starts in the state $\ket{u_r}$, then the algorithm leaves $\mathsf A$ in this state,
  and $\widehat M\ge M-h/2$ with probability at least $1-\delta$.
  If moreover $\dist_{\T}\rbra{\phi_j,0}\ge2\gamma$ for every $j$, then also $\abs{\widehat M-M}\le h/2$ with probability at least $1-\delta$.
  The same guarantees hold conditional on every preceding history of positive probability, provided that the registers start in the states specified above.
\end{lemma}

\begin{proof}
We adapt the fuzzy interval search of \cite{LinTong2020,DongLinTong2022}. The construction below does not depend on $r$. Fix $r$. We construct nested random intervals $\interval{L_k}{H_k}$ with $L_0=\pi$ and $H_0=2\pi$.
Their widths and midpoints are $w_k=H_k-L_k$ and $t_k=\rbra{L_k+H_k}/2$, respectively.
Whenever $w_k>h$, let
\begin{align*}
  s_k=\frac{\min\cbra{\gamma,w_k/8}}{20},\qquad
  \delta_k=\frac{3\delta h}{8w_k}.
\end{align*}
Let $b_k=\textnormal{\textsc{Search}}\rbra{F^\sigma_{s_k,t_k},\delta_k}$ be the output of \cref{lem:neutral-search}.
If $b_k=1$, the next interval is $\interval{L_k+3w_k/8}{H_k}$, and otherwise it is $\interval{L_k}{H_k-3w_k/8}$.
Both possible intervals have width $\rho w_k$, where $\rho=5/8$, so $w_k=\pi\rho^k$.
Their overlap is $\interval{t_k-w_k/8}{t_k+w_k/8}$.
Denote the stopping index by
\begin{align*}
  K=\min\cbra*{k\in\Z_{\ge0}:\pi\rho^k\le h}
\end{align*}
and estimate by
\begin{align*}
  M=\frac{L_K+H_K}{2}.
\end{align*}
Each routing circuit returns the reference register $\mathsf A$ to $\ket{u_r}$ by \cref{eq:routing}, and all other gates leave $\mathsf A$ untouched.
Thus every search run uses the same reference index $r$ and hence the same phases $\phi_j$.
The guarantees for $b_k$ hold conditional on every preceding history of positive probability.

\emph{Failure probability.}
Since $h<\pi$, we have $K\ge1$, and the definition of $K$ gives $h<\pi\rho^{K-1}\le h/\rho$.
Since $3/8=1-\rho$ and $w_k=\pi\rho^k$,
\begin{align*}
  \sum_{k=0}^{K-1}\delta_k
  =\frac{\rbra{1-\rho}\delta h}{\pi\rho^{K-1}}\sum_{m=0}^{K-1}\rho^m
  <\delta.
\end{align*}
In particular, each $\delta_k$ lies in $\interval[open]{0}{1/3}$.

\emph{One-sided accuracy.}
Suppose $M$ lies above the overlap.
By \cref{lem:phase-estimation}, for each label the phase estimate has circular error at most
$s_k<\min\cbra{\gamma,w_k/8}$ with probability at least $19/20$. We call such an estimate accurate.
If a phase equals zero, an accurate estimate of that phase is accepted near the cut.
Otherwise, take $j$ with $\phi_j=M$. Every accurate estimate of $\phi_j$ is either at least $M-s_k>t_k$
or, after wrapping past $2\pi$, within distance $s_k<\gamma$ of the cut.
Thus, by \cref{lem:cut-verifier}, some label has acceptance probability at least $19/20$.
By \cref{lem:neutral-search}, $b_k=0$ with probability at most $\delta_k$.
The upper endpoint can drop below $M$ only in this case with $b_k=0$.
Since $H_0=2\pi\ge M$ and $\sum_k\delta_k<\delta$, a conditional union bound shows that
$H_k\ge M$ for all $0\le k\le K$ with probability greater than $1-\delta$.
On this event, $\widehat M=H_K-w_K/2\ge M-h/2$.

\emph{Two-sided accuracy away from the cut.}
Now suppose $\dist_{\T}\rbra{\phi_j,0}\ge2\gamma$ for every $j$, so $M=\max_j\phi_j$.
Since $\dist_{\T}\rbra{\phi_j,0}\ge2\gamma>s_k$, an accurate estimate of $\phi_j$ does not wrap around the cut, so it has absolute error at most $s_k$ and lies at distance more than $\gamma$ from the cut.
If $M$ lies below the overlap, every such estimate lies below $t_k$, so by \cref{lem:cut-verifier} every label has acceptance probability at most $1/20$,
and \cref{lem:neutral-search} gives $b_k=1$ with probability at most $\delta_k$.
If $M$ lies in the overlap, both possible intervals contain $M$.
Otherwise, $M$ can be lost only through $b_k=0$ when $M$ lies above the overlap, or through $b_k=1$ when it lies below.
Consequently, whenever $M\in\interval{L_k}{H_k}$,
the probability that $M\notin\interval{L_{k+1}}{H_{k+1}}$ is at most $\delta_k$.
Since $M\in\interval{L_0}{H_0}$ and $\sum_k\delta_k<\delta$, it follows that
$M\in\interval{L_k}{H_k}$ for all $0\le k\le K$ with probability greater than $1-\delta$.
On this event, $\abs{\widehat M-M}\le w_K/2\le h/2$.

\emph{Query complexity.}
Each application of $F^\sigma_{s_k,t_k}$ or its inverse uses $O\rbra{1/s_k}$ queries
by \cref{lem:cut-verifier}.
The decision $b_k$ therefore requires
$O\rbra{\sqrt d\,s_k^{-1}\log\rbra{1/\delta_k}}$ queries by \cref{lem:neutral-search}.
For $m=K-1-k$, the identity $w_k=\pi\rho^{K-1}\rho^{-m}$ and the bounds $h<\pi\rho^{K-1}\le h/\rho$ give $s_k^{-1}=O\rbra*{\gamma^{-1}+h^{-1}\rho^m}$, and $\log\rbra*{1/\delta_k}=O\rbra*{\log\rbra*{1/\delta}+m+1}$.
Summing over $k$ and using $\sum_{m\ge0}\rbra{m+1}\rho^m=O\rbra{1}$, the total number of queries is
\begin{align*}
  O\rbra*{\sqrt d\rbra*{
    \frac{K\log\rbra{1/\delta}+K^2}{\gamma}
    +\frac{\log\rbra{1/\delta}}{h}}}.
\end{align*}
Since $h=\eps/4$, we have $K=O\rbra{\log\rbra{1/\eps}}$.
Substituting $\gamma=\sqrt\eps/4$ and $h=\eps/4$, and using that $\sqrt\eps\log\rbra{1/\eps}$ and $\sqrt\eps\log^2\rbra{1/\eps}$ are bounded, this cost is
$O\rbra{\sqrt d\log\rbra{1/\delta}/\eps}$.
The widths and the parameters $s_k$ and $\delta_k$ do not depend on the outcomes, so this bound holds on every execution branch.
\end{proof}

\subsection{Diamond distance estimator}
\label{subsec:complete-estimator}
The complete estimator is formally given in \Cref{alg:estimator}.

\begin{algorithm}[!htb]
  \caption{Diamond distance estimator}
  \label{alg:estimator}
  \begin{algorithmic}[1]
    \Require query access to $U$, $V$, $U^\dag$, and $V^\dag$ for $U,V\in\mathbb U\rbra{d}$, where $d\ge2$, and $\eps\in\interval[open left]{0}{1/4}$
    \Ensure an estimate of $\Abs{\cU-\cV}_\diamond$
    \State $\gamma\gets\sqrt\eps/4$, $h\gets\eps/4$, $\delta\gets1/6$
    \State $K\gets\left\lceil\log_{8/5}\rbra{4\pi/\eps}\right\rceil$
    \State Sample $i\in\sbra{d}$ uniformly at random and prepare $\mathsf A$ in $\ket{i}$
      \Comment{$\mathsf A$ is never measured}
    \For{$\sigma=+1$ and then $\sigma=-1$}
      \State $L\gets\pi$, $H\gets2\pi$
      \For{$k=0,\ldots,K-1$}
        \State $w\gets H-L$, $t\gets\rbra{L+H}/2$, $s\gets\min\cbra{\gamma,w/8}/20$, $\delta'\gets3\delta h/\rbra{8w}$
        \State Construct the verifier $F^\sigma_{s,t}$ of \cref{lem:cut-verifier}
        \State $b\gets\textsc{Search}\rbra{F^\sigma_{s,t},\delta'}$ \Comment{\cref{lem:neutral-search}}
        \If{$b=1$}
          \State $L\gets L+3w/8$
        \Else
          \State $H\gets H-3w/8$
        \EndIf
      \EndFor
      \State $\widehat M_\sigma\gets\rbra{L+H}/2$
        \Comment{$\widehat M_+$ for $\sigma=+1$, $\widehat M_-$ for $\sigma=-1$}
    \EndFor
    \State \Return $2\sin\rbra[\big]{\tfrac12\min\cbra{\pi,\widehat M_++\widehat M_--2\pi}}$
  \end{algorithmic}
\end{algorithm}

\begin{proof}[Proof of \cref{thm:main}]
We analyze \cref{alg:estimator} with the parameters set in its first two lines.
For $\sigma=+1$ and $\sigma=-1$, its inner loop is the algorithm of \cref{lem:fuzzy-max} with the retained register $\mathsf A$,
and returns $\widehat M_+$ and $\widehat M_-$, respectively.
Let $\widehat D$ be its output.
Both estimates $\widehat M_+$ and $\widehat M_-$ lie in $\interval{\pi}{2\pi}$, so $\widehat D\in\interval{0}{2}$ on every branch.

Fix a reference index $r\in\sbra{d}$, write $M_+$ and $M_-$ for the values of $M$ for $\sigma=+1$ and $\sigma=-1$,
and set $\phi_j=\phi_{j,r}$, the phases for $\sigma=+1$.
Let
\begin{align*}
  a\coloneqq\min_j\dist_{\T}\rbra*{\phi_j,0},
  \qquad D\coloneqq\Abs*{\cU-\cV}_\diamond.
\end{align*}
Replacing $\sigma=+1$ by $\sigma=-1$ does not change $a$.
We first prove accuracy when both estimates satisfy the applicable guarantee of \cref{lem:fuzzy-max},
and then bound the probability of this event.

\emph{Case $a\ge2\gamma$.}
In this case no phase lies at the cut, so the phases for $\sigma=-1$ are $2\pi-\phi_j$.
Thus $\widehat M_+$ estimates $\max_j\phi_j$ and $\widehat M_-$ estimates $2\pi-\min_j\phi_j$.
The two-sided guarantees give
\begin{align*}
  \abs*{\widehat M_++\widehat M_--2\pi
    -\rbra*{\max_j\phi_j-\min_j\phi_j}}\le h.
\end{align*}
Truncating at $\pi$ and applying $2\sin\rbra{x/2}$ on $\interval{0}{\pi}$ are $1$-Lipschitz operations.
By \cref{lem:relative-width,prop:width-formula}, the final error is at most $h=\eps/4$.

\emph{Case $a<2\gamma$.}
By the definition of $a$, some eigenphase $\theta_j$ satisfies $\dist_{\T}\rbra{\theta_j,\theta_r}=\pi-a$.
Hence $\ell\rbra{W}\ge\pi-a$, and since $\pi-a\le\pi$, also $\omega\rbra{W}\ge\pi-a$.
By \cref{prop:width-formula} and $a<2\gamma$,
\begin{align*}
  D=2\sin\rbra*{\frac{\omega\rbra{W}}{2}}\ge2\sin\rbra*{\frac{\pi-a}{2}}=2\cos\rbra*{\frac a2}\ge2\cos\gamma.
\end{align*}
If $a>0$, one of $M_+$ and $M_-$ is at least $2\pi-a$.
If $a=0$, both equal $2\pi$ under the convention for $M$.
Both $M_+$ and $M_-$ are at least $\pi$, because $\phi_r=\pi$.
Applying the one-sided guarantees in the two directions gives
\begin{align*}
  \widehat M_++\widehat M_--2\pi\ge\pi-a-h\ge\pi-2\gamma-h.
\end{align*}
Since $2\gamma+h<\pi$, truncation at $\pi$ preserves this lower bound and gives
\begin{align*}
  \widehat D\ge2\cos\rbra*{\gamma+h/2}.
\end{align*}
Both $D$ and $\widehat D$ lie in $\interval{2\cos\rbra{\gamma+h/2}}{2}$, so
\begin{align*}
  \abs*{\widehat D-D}
  \le2-2\cos\rbra*{\gamma+h/2}
  \le\rbra*{\gamma+h/2}^2
  \le\eps.
\end{align*}

\emph{Success probability.}
For fixed $r$, each estimate violates its required guarantee with probability at most $\delta$ by \cref{lem:fuzzy-max}.
Thus both guarantees hold with probability at least $1-2\delta=2/3$.

Averaged over its random initialization, $\mathsf A$ is in the state $\frac1d\sum_{r=0}^{d-1}\ketbra{u_r}{u_r}$.
By linearity, the output distribution is the uniform mixture over fixed reference indices.
Each component estimates the same distance with success probability at least $2/3$,
which therefore also holds for $\widehat D$.

\emph{Query complexity.}
With $\delta=1/6$, each of $\widehat M_+$ and $\widehat M_-$ requires $O\rbra{\sqrt d/\eps}$ queries by \cref{lem:fuzzy-max}.
The same bound therefore holds for $\widehat D$ on every branch.
\end{proof}

\paragraph{Acknowledgements.}
J.~B.\ was supported by the Quantum Advantage Turbo Charger (QATCH) programme, funded by UKRI under Grant No.~\mbox{UKRI4257}.
Q.~W.\ was supported by the Startup Funding from Shanghai Jiao Tong University. 

\paragraph{AI use statement.}
The authors used Claude (Anthropic) and Codex (OpenAI), AI assistants, to proofread the manuscript,
check the consistency of notation and cross-references, compare citation details with the cited sources,
and draft revisions of the exposition, the \LaTeX{} source, and the figures.
The authors checked all mathematical statements and proofs and take full responsibility for the content of this paper.

\bibliographystyle{alphaurl}
\bibliography{references}

@inproceedings{AharonovKitaevNisan1998,
  author        = {Aharonov, Dorit and Kitaev, Alexei Y. and Nisan, Noam},
  title         = {Quantum Circuits with Mixed States},
  booktitle     = {Proceedings of the 30th Annual ACM Symposium on Theory of Computing},
  pages         = {20--30},
  year          = {1998},
  doi           = {10.1145/276698.276708},
}

@article{Acin2001,
  author        = {Ac{\'i}n, Antonio},
  title         = {Statistical Distinguishability between Unitary Operations},
  journal       = {Physical Review Letters},
  volume        = {87},
  number        = {17},
  pages         = {177901},
  year          = {2001},
  doi           = {10.1103/PhysRevLett.87.177901},
}

@inproceedings{ChenWangZhang2026,
  author        = {Chen, Kean and Wang, Qisheng and Zhang, Zhicheng},
  title         = {Strict Hierarchy for Quantum Channel Certification to Unitary},
  booktitle     = {Proceedings of the 53rd International Colloquium on Automata, Languages, and Programming},
  pages         = {59:1--59:12},
  year          = {2026},
  doi           = {10.4230/LIPIcs.ICALP.2026.59},
}

@article{DongLinTong2022,
  author        = {Dong, Yulong and Lin, Lin and Tong, Yu},
  title         = {Ground-State Preparation and Energy Estimation on Early Fault-Tolerant Quantum Computers via Quantum Eigenvalue Transformation of Unitary Matrices},
  journal       = {PRX Quantum},
  volume        = {3},
  number        = {4},
  pages         = {040305},
  year          = {2022},
  doi           = {10.1103/PRXQuantum.3.040305},
}

@inproceedings{HaahKothariODonnellTang2023,
  author        = {Haah, Jeongwan and Kothari, Robin and O'Donnell, Ryan and Tang, Ewin},
  title         = {Query-Optimal Estimation of Unitary Channels in Diamond Distance},
  booktitle     = {Proceedings of the 64th IEEE Annual Symposium on Foundations of Computer Science},
  pages         = {363--390},
  year          = {2023},
  doi           = {10.1109/FOCS57990.2023.00028},
}

@inproceedings{HoyerMoscaDeWolf2003,
  author        = {H{\o}yer, Peter and Mosca, Michele and de Wolf, Ronald},
  title         = {Quantum Search on Bounded-Error Inputs},
  booktitle     = {Proceedings of the 30th International Colloquium on Automata, Languages, and Programming},
  volume        = {2719},
  pages         = {291--299},
  year          = {2003},
  doi           = {10.1007/3-540-45061-0_25},
}

@misc{JefferyWitteveen2026,
  title  = {Optimal Quantum Algorithm for Ground-State Energy Estimation with a Guiding State},
  author = {Jeffery, Stacey and Witteveen, Freek},
  howpublished   = {ArXiv preprints},
  eprint = {2608.24494},
  year   = {2026}
}

@article{JeonOh2026,
  author        = {Jeon, Sangwoo and Oh, Changhun},
  title         = {On the Query Complexity of Unitary Channel Certification},
  journal       = {npj Quantum Information},
  volume        = {12},
  pages         = {2},
  year          = {2026},
  doi           = {10.1038/s41534-025-01135-5},
}

@misc{Kitaev1995,
  title  = {Quantum Measurements and the Abelian Stabilizer Problem},
  author = {Kitaev, Alexei Yu.},
  howpublished   = {ArXiv preprints},
  eprint = {quant-ph/9511026},
  year   = {1995}
}

@inproceedings{KociumakaNoglerWellnitz2025,
  author        = {Kociumaka, Tomasz and Nogler, Jakob and Wellnitz, Philip},
  title         = {Near-Optimal-Time Quantum Algorithms for Approximate Pattern Matching},
  booktitle     = {Proceedings of the 2025 Annual ACM-SIAM Symposium on Discrete Algorithms},
  pages         = {517--534},
  year          = {2025},
  doi           = {10.1137/1.9781611978322.15},
}

@article{LinTong2020,
  author        = {Lin, Lin and Tong, Yu},
  title         = {Near-Optimal Ground State Preparation},
  journal       = {Quantum},
  volume        = {4},
  pages         = {372},
  year          = {2020},
  doi           = {10.22331/q-2020-12-14-372},
}

@article{MandeDeWolf2026,
  author        = {Mande, Nikhil S. and de Wolf, Ronald},
  title         = {Tight Bounds for Quantum Phase Estimation and Related Problems},
  journal       = {Quantum},
  volume        = {10},
  pages         = {2140},
  year          = {2026},
  doi           = {10.22331/q-2026-06-15-2140},
}

@article{vanApeldoornGilyenGriblingDeWolf2020,
  author        = {van Apeldoorn, Joran and Gily{\'e}n, Andr{\'a}s and Gribling, Sander and de Wolf, Ronald},
  title         = {Quantum {SDP}-Solvers: Better Upper and Lower Bounds},
  journal       = {Quantum},
  volume        = {4},
  pages         = {230},
  year          = {2020},
  doi           = {10.22331/q-2020-02-14-230},
}

@book{Watrous2018,
  author    = {Watrous, John},
  title     = {The Theory of Quantum Information},
  publisher = {Cambridge University Press},
  year      = {2018},
  doi       = {10.1017/9781316848142},
}

@article{KLL+17,
    author = {Kimmel, Shelby and Lin, Cedric Yen-Yu and Low, Guang Hao and Ozols, Maris and Yoder, Theodore J.},
    title = {Hamiltonian simulation with optimal sample complexity},
    journal = {npj Quantum Information},
    volume = {3},
    pages = {13},
    doi = {10.1038/s41534-017-0013-7},
    year = {2017}
}

@misc{CGO+26,
  title  = {Quantum channel tomography: optimal bounds and a {Heisenberg}-to-classical phase transition},
  author = {Chen, Kean and Girardi, Filippo and Oufkir, Aadil and Yu, Nengkun and Zhang, Zhicheng},
  howpublished   = {ArXiv preprints},
  eprint = {2604.17369},
  year   = {2026}
}

@article{YRC20,
    author = {Yang, Yuxiang and Renner, Renato and Chiribella, Giulio},
    title = {Optimal Universal Programming of Unitary Gates},
    journal = {Physical Review Letters},
    volume = {125},
    number = {21},
    pages = {210501},
    doi = {10.1103/PhysRevLett.125.210501},
    year = {2020}
}

@inproceedings{FFGO23,
    author = {Fawzi, Omar and Flammarion, Nicolas and Garivier, Aur{\'{e}}lien and Oufkir, Aadil},
    title = {Quantum channel certification with incoherent measurements},
    booktitle = {Proceedings of the 36th Conference on Learning Theory},
    pages = {1822--1884},
    url = {https://proceedings.mlr.press/v195/fawzi23a.html},
    year = {2023}
}

@article{Wat09,
    author = {Watrous, John},
    title = {Semidefinite Programs for Completely Bounded Norms},
    journal = {Theory of Computing},
    volume = {5},
    number = {},
    pages = {217--238},
    doi = {10.4086/toc.2009.v005a011},
    year = {2009}
}

@article{BATS10,
    author = {Ben-Aroya, Avraham and Ta-Shma, Amnon},
    title = {On the complexity of approximating the diamond norm},
    journal = {Quantum Information and Computation},
    volume = {10},
    number = {1},
    pages = {77--86},
    doi = {10.26421/QIC10.1-2-6},
    year = {2010}
}

@article{KR21,
    author = {Kliesch, Martin and Roth, Ingo},
    title = {Theory of Quantum System Certification},
    journal = {PRX Quantum},
    volume = {2},
    number = {1},
    pages = {010201},
    doi = {10.1103/PRXQuantum.2.010201},
    year = {2021}
}

@misc{JW09,
  title  = {Non-identity check remains {QMA}-complete for short circuits},
  author = {Ji, Zhengfeng and Wu, Xiaodi},
  howpublished   = {ArXiv preprints},
  eprint = {0906.5416},
  year   = {2009}
}

@misc{ZhengWangYuXuZhang2025,
  title  = {Optimal Distributed Similarity Estimation of Quantum Channels},
  author = {Zheng, Congcong and Wang, Kun and Yu, Xutao and Xu, Ping and Zhang, Zaichen},
  howpublished   = {ArXiv preprints},
  eprint = {2512.10465},
  year   = {2025}
}

@article{Cam19,
    author = {Campbell, Earl},
    title = {Random Compiler for Fast {Hamiltonian} Simulation},
    journal = {Physical Review Letters},
    volume = {123},
    number = {7},
    pages = {070503},
    doi = {10.1103/PhysRevLett.123.070503},
    year = {2019}
}

@article{BCS+20,
    author = {Berry, Dominic W. and Childs, Andrew M. and Su, Yuan and Wang, Xin and Wiebe, Nathan},
    title = {Time-dependent Hamiltonian simulation with {$L^1$-norm} scaling},
    journal = {Quantum},
    volume = {4},
    number = {},
    pages = {254},
    doi = {10.22331/q-2020-04-20-254},
    year = {2020}
}

\end{document}